\documentclass[preprint2]{aastex}

\newcommand{\xmm}{{\it XMM-Newton}\ }
\newcommand{\asca}{{\it ASCA}\ }
\newcommand{\chandra}{{\it Chandra}\ }
\newcommand{\sax}{{\it Beppo-SAX}\ }
\newcommand{\suzaku}{{\it Suzaku}\ }

\slugcomment{Draft \today}

\shorttitle{Perseus cluster}
\shortauthors{Tamura et al.}

\begin{document}

\title{X-ray Spectroscopy of the Core of the Perseus Cluster with \suzaku:
  Elemental Abundances in the Intracluster Medium
}

\author{T. Tamura, Y. Maeda, K. Mitsuda}
\affil{Institute of Space and Astronautical Science,
Japan Aerospace Exploration Agency,\\
3-1-1 Yoshinodai, Sagamihara, Kanagawa 229-8510, Japan
}

\author{A. C. Fabian, J. S. Sanders}
\affil{Institute of Astronomy, Madingley Road, Cambridge CB3 0HA, UK}

\author{A. Furuzawa}
\affil{Department of Particle and Astrophysics, Nagoya University, Furo-cho, Chikusa-ku, Nagoya 464-8602, Japan}

\author{J. P. Hughes}
\affil{Department of Physics and Astronomy, Rutgers University, Piscataway, NJ 08854-8019, USA}

\author{R. Iizuka}
\affil{Department of Physics, Faculty of Science and Engineering, Chuo University, 1-13-27 Kasuga, Bunkyo-ku, Tokyo 112-8551, Japan}

\author{K. Matsushita}
\affil{Department of Physics, Tokyo University of Science, 1-3 Kagurazaka, Shinjyuku-ku, Tokyo 162-8601, Japan}

\and

\author{T. Tamagawa}
\affil{RIKEN (Institute of Physical and Chemical Research), 2-1 Hirosawa, Wako, Saitama 351-0198, Japan}

\begin{abstract}
The results from \suzaku observations of the central region of the Perseus cluster are presented.
Deep exposures with the X-ray Imaging Spectrometer provide high quality X-ray spectra from the intracluster medium.
X-ray lines from helium-like Cr and Mn have been detected significantly for the first time in clusters.
In addition, elemental abundances of Ne, Mg, Si, S, Ar, Ca, Fe, and Ni are accurately measured within $10'$ (or 220~kpc) from the cluster center.
The relative abundance ratios are found to be within a range of $0.8-1.5$ times the solar value.
These abundance ratios are compared with previous measurements, 
those in extremely metal-poor stars in the Galaxy, and
theoretical models.

\end{abstract}


\section{INTRODUCTION}
Heavy elements (or metals) in the universe are created in stars and supernovae (SNe).
These metals enrich the interstellar medium and are recycled into generations of stars.
Some part of these should have been transported into intergalactic space via galactic winds or ram pressure stripping.
Indeed, the intracluster medium (ICM) contains an amount of metals comparable to the total amount of metals found in galaxies \citep[e.g.][]{tsuru91}.
The distribution of these metals can be measured exclusively by X-ray spectroscopy.
This constrains the total number of SNe in galaxies and the efficiency of the mass transport from galaxies into intergalactic space.
Moreover, the abundance ratios among elements (e.g.  Si/Fe ratio) can be used to identify their origins \citep[e.g.][]{renzini93}.

\asca and \sax observations provided the first systematic measurements of Si, S and Fe abundances and their spatial distributions \citep[e.g.][]{mush96}. 
\xmm has allowed more detailed studies including the O abundance determination \citep[e.g.][]{tamura04}.
Recent \suzaku data have further improved measurements of elements such as O and Mg \citep[e.g.][]{matsushita07}.
Mostly, observed abundance patterns are explained by some combinations of origins from type Ia and type II SNe \citep[e.g.][]{fuka98}.
Note that the nature of these SNe and hence their total metal products have not completely been understood.
In addition, some authors claim that the observed metal abundances require an alternative source such as hypernovae in early generations of stars \citep{loe01, bau05}.
However, measurements are still limited to abundant elements and even for those elements statistical and systematic errors are large.
See \citet{Boehringer09} for a recent review on X-ray spectroscopy of clusters.

Here we present \suzaku XIS (X-ray Imaging Spectrometer) spectroscopy of the central region of the Perseus cluster of galaxies, which is the brightest cluster in X-rays.
This is a calibration target of the XIS 
and hence has been observed regularly.
We used all available data from several exposures.
The deep exposure of this unique object along with the high sensitivity of the XIS 
provides one of the best quality X-ray spectra from clusters.
We have detected X-ray line emission from ionized Cr and Mn for the first time from an extragalactic source.
Furthermore we accurately measured elemental abundances of Ne, Mg, Si, S, Ar, Ca, Fe, and Ni. 

Throughout this paper,
we assume cosmological parameters as follows; $H_0 = 70$ km s$^{-1}$Mpc$^{-1}$, 
$\Omega_\mathrm{m} = 0.3$, and $\Omega_\mathrm{\Lambda} = 0.7$.
% http://heasarc.nasa.gov/xanadu/xspec/manual/XScosmo.html
At the cluster redshift of 0.0183, 
% from Sanders 2007
one arc-minute corresponds to 22.2~kpc at the cluster distance.
We use the 90\% confidence level.

\section{OBSERVATIONS}\label{sect:obs}
As a calibration target, the Perseus cluster has been observed twice a year in various operation modes.
Here we use the data obtained from February 2006 to February 2009 in the normal window mode.
Sequence numbers of the data are 
800010010, 101012010, 101012020, 102011010, 102012010, 103004010, 103005010, and 103004020.
The central galaxy, NGC~1275, has been located at the CCD center.
The detector covers a central $17' \times 17'$ square region.
The XIS~2 sensor was available only until 2006.
% i02111@klm:/home/cc/i02111/KLM/suzaku/perseus/xis/0423-1>Exp.csh
% /home/cc/i02111/KLM/suzaku/perseus/xis/0423-1//scion/xisFIr0.5_2_sum.pi
% EXPOSURE=         5.803612E+05 / exposure (in seconds)
% /home/cc/i02111/KLM/suzaku/perseus/xis/0423-1///scion/xis1r0.5_2_sum.pi
% EXPOSURE=         2.653599E+05 / exposure (in seconds)
% /home/cc/i02111/KLM/suzaku/perseus/xis/0423-1//scioff/xisFIr0.5_2_sum12.pi.grp11
% EXPOSURE=         2.996675E+05 /  Exposure time
The cumulative exposure time in the spaced-row charge injection {\it off} mode is
300~ks times sensors (XIS~0, XIS~2, and XIS~3; FI)\footnote{BI data in spaced-row charge injection off are not used.}.
Those in the spaced-row charge injection {\it on} mode are
580~ks times sensors (FI) and 265~ks (XIS~1; BI).
Detailed descriptions of the \suzaku observatory, the XIS instrument, 
and the X-ray telescope  are found in 
\citet{mitsuda07}, \citet{koyama07}, \citet{serlemitsos07}, respectively.

\section{ANALYSIS AND RESULTS}
\subsection{Method}
We started the analysis from archived cleaned event files, 
which were filtered with the standard selection criteria.
We used the latest calibration file as of May 2009.
We examined light curves excluding the central bright region events ($R<6'$) for stable-background periods.
There was no flaring event in all data.
% (nxb)
The instrumental (Non-X-ray) background was estimated using the night earth observation database and the software {\tt xisnxbgen} \citep{tawa08}.
% cxb
The cosmic X-ray background was estimated to be well below 1\% of the source over almost entire enegy band.
We therefore ignore this cosmic background in the following analysis. % unless stated otherwise .

%\subsection{Spectral Extraction and Fitting method}
We extracted spectra from three annulus regions with boundary radii of $0'.5$, $2'$, $4'.0$, and $10'$ centered on the X-ray maximum.
The central region ($R<0'.5$) is excluded to avoid the contribution from the central galaxy.

% (arf, rmf)
We prepared an X-ray telescope response function for each spectrum using the XIS ARF builder software {\tt xissimarfgen} \citep{ishisaki07},
assuming an ICM surface brightness distribution based on an \xmm public image of the cluster
as a source image.
We simulated events so that the relative error of the response at each energy bin is smaller than 0.25\% (using a parameter {\tt accuracy}).
To made the XIS energy response function, we used the software {\tt xisrmfgen}.

To describe the thermal emission from a collisional ionization equilibrium plasma, 
we use the APEC model \citep{smith01} with the solar metal abundances
taken from \citet{ag89}.
% http://heasarc.gsfc.nasa.gov/docs/xanadu/xspec/manual/XSabund.html
% O/H = 8.51e-04, Fe/H=3.63e-04
The Galactic absorption is described by the photo-electric absorption of Wisconsin cross-sections (wabs model in XSPEC)
with the reported neutral hydrogen column density of $1.5\times 10^{-21}$cm$^{-2}$ toward the cluster.
%http://heasarc.nasa.gov/xanadu/xspec/manual/XSmodelWabs.html

\subsection{Spectral Fitting and Elemental Abundances}
We fit the data to constrain the thermal nature and metal abundances.
To describe the complex thermal structure of the central cooling core, 
projection effect, and photon mixing due to the point spread function,
we introduce two thermal components.
Temperatures and iron abundances of the two components are made independently free.
On the contrary, other abundances of Ne, Mg, Si, S, Ar, Ca, and Ni are coupled between the two components, 
e.g. Ne$_1$ $=$ Ne$_2$. 
A power-law component with a fixed photon index of 1.8 is also included.
This describes a combination of scattering of the central galaxy emission 
and the possible distributed hard X-ray emission suggested by \citet{sanders04}.
In some spectra we found line emission at the position of He-like Cr.
Because there is no line from Cr included in the APEC model, here, we add a Gaussian component for the Cr line into the fitting model.
A detailed analysis of this line is given in the next subsection.

To avoid calibration errors, we ignore energy ranges around the Si K edge ($1.6-1.95$~keV) and the Au M edge ($2.2-2.4$~keV).
To compensate errors of the energy-PI relation, redshifts of the two APEC components are allowed to be different.
In $0'.5-2'-4'$ regions, we use the FI and BI spectra simultaneously. 
In the $4'-10'$ region, we use only the FI spectrum, since the BI spectrum has a relatively low signal to background ratio at higher energies, $>5$~keV.

\begin{figure*}
\begin{center}
\includegraphics[scale=.28]{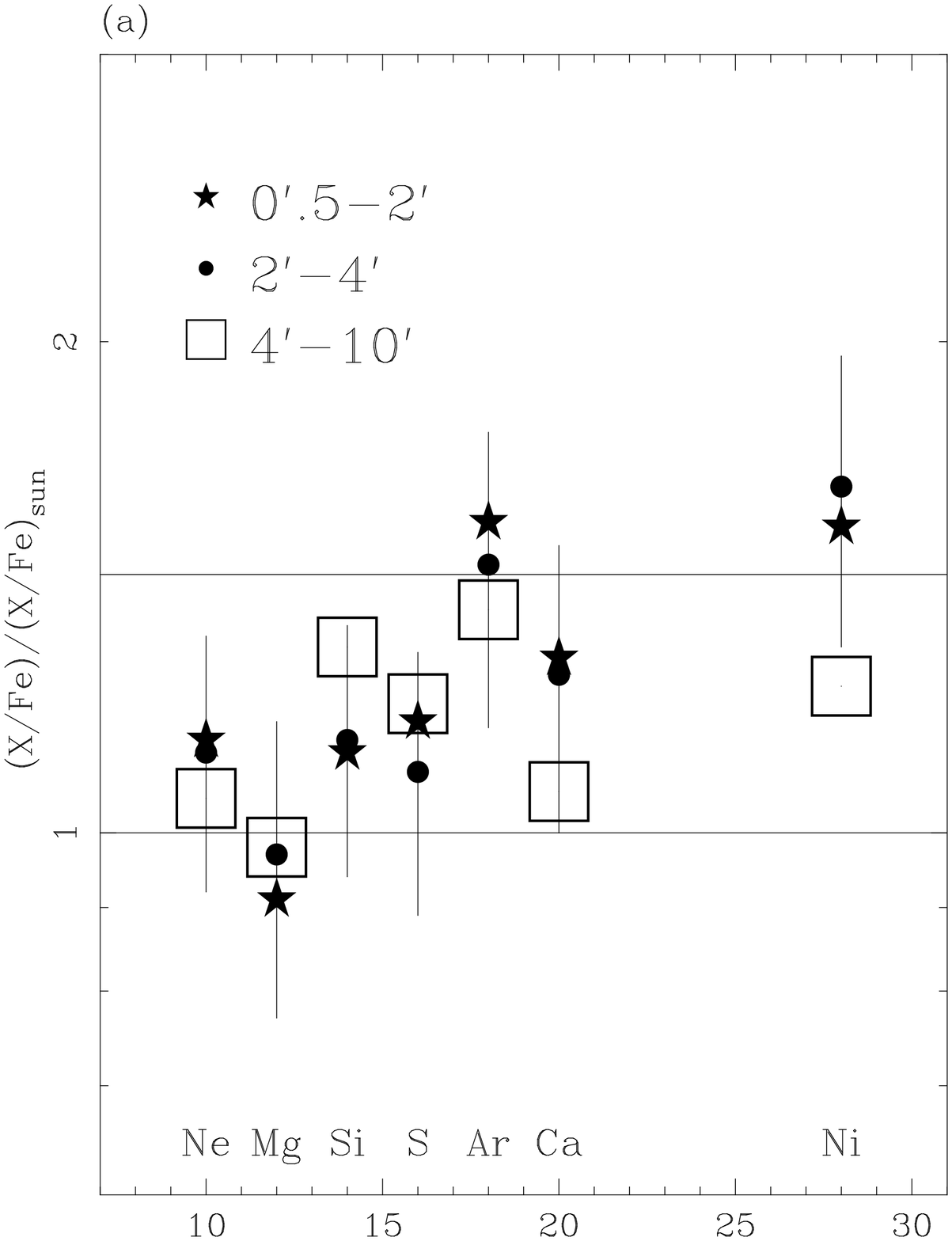}
\includegraphics[scale=.28]{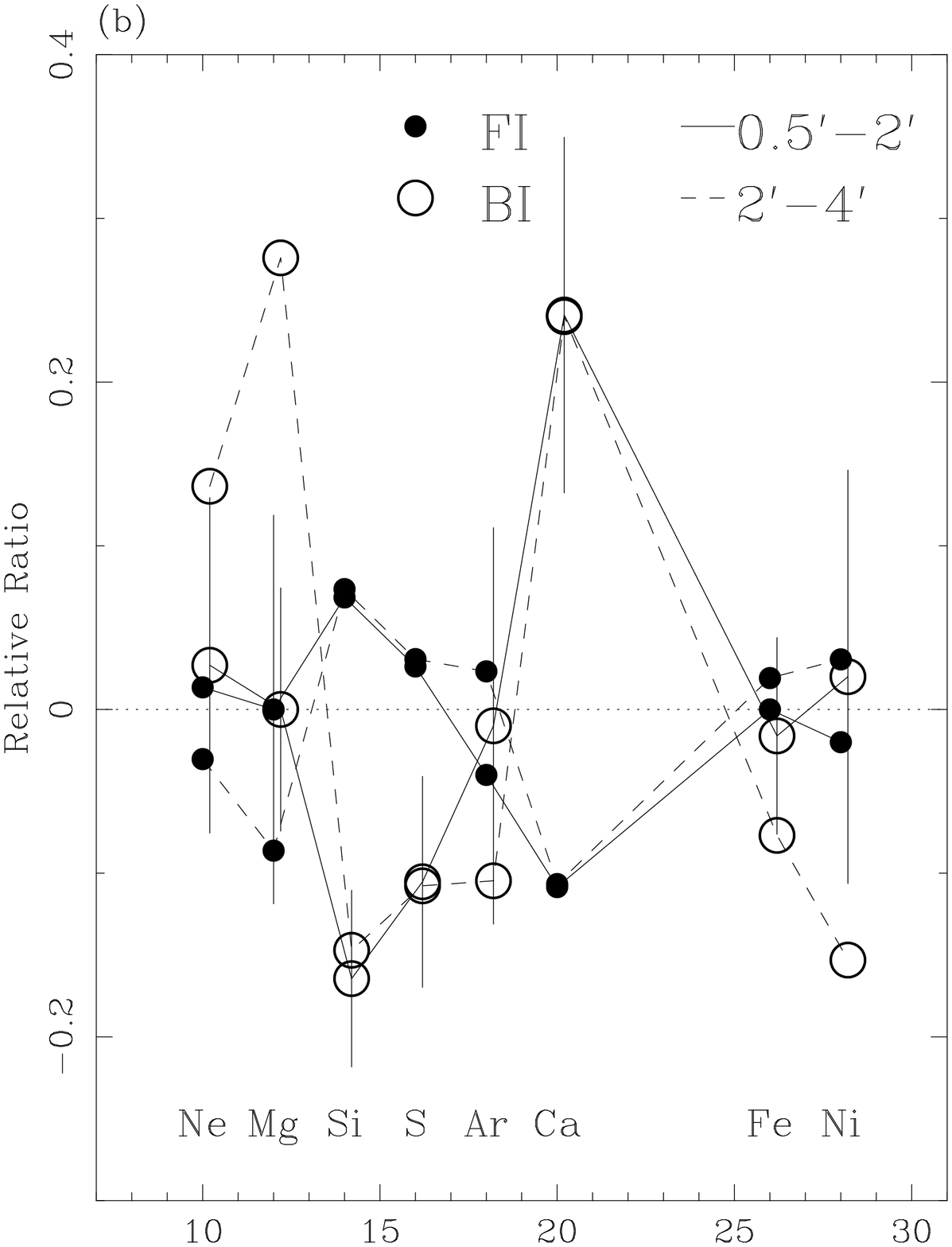}
\end{center}
\begin{center}
\includegraphics[scale=.28]{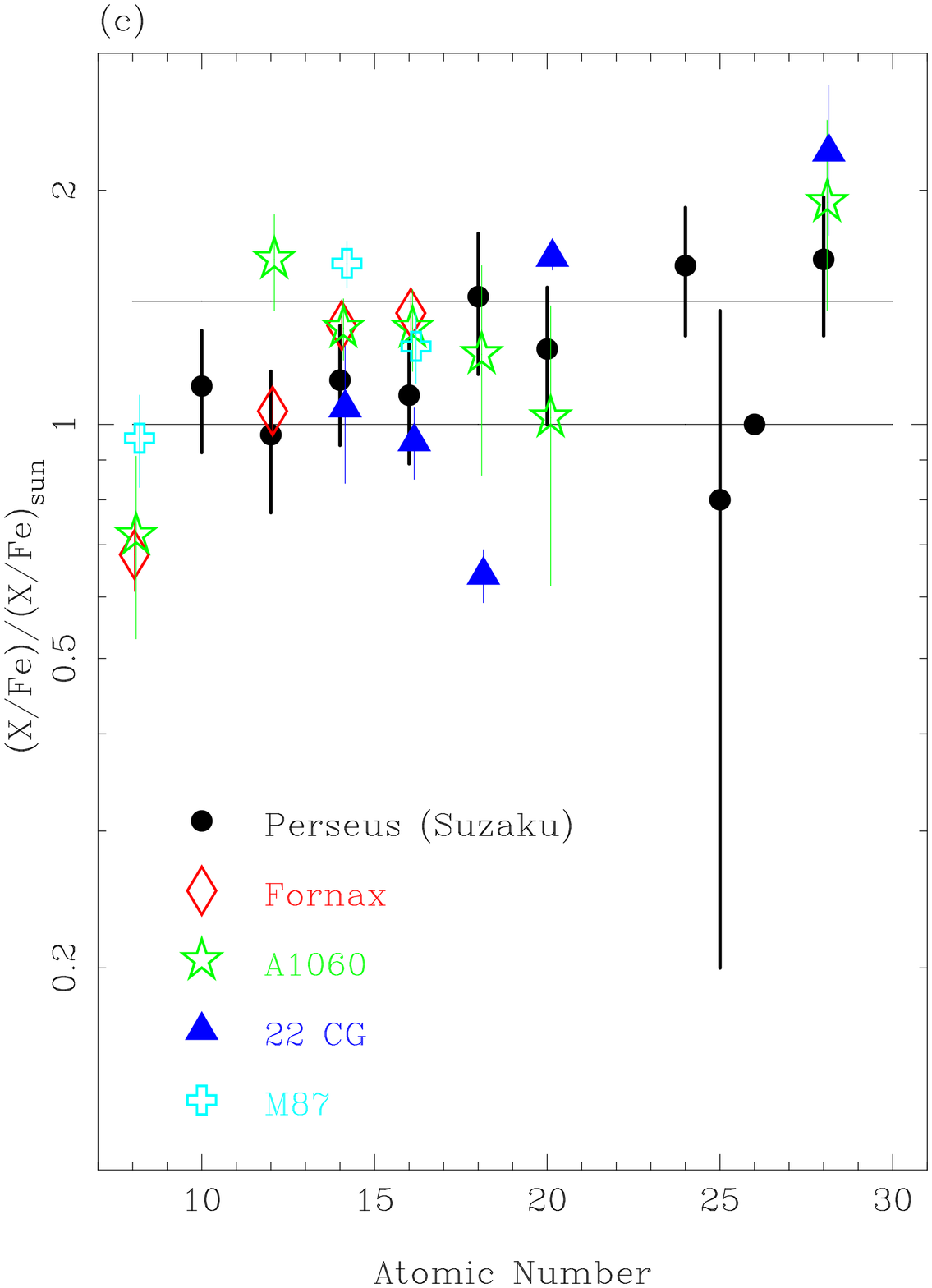}
\includegraphics[scale=.28]{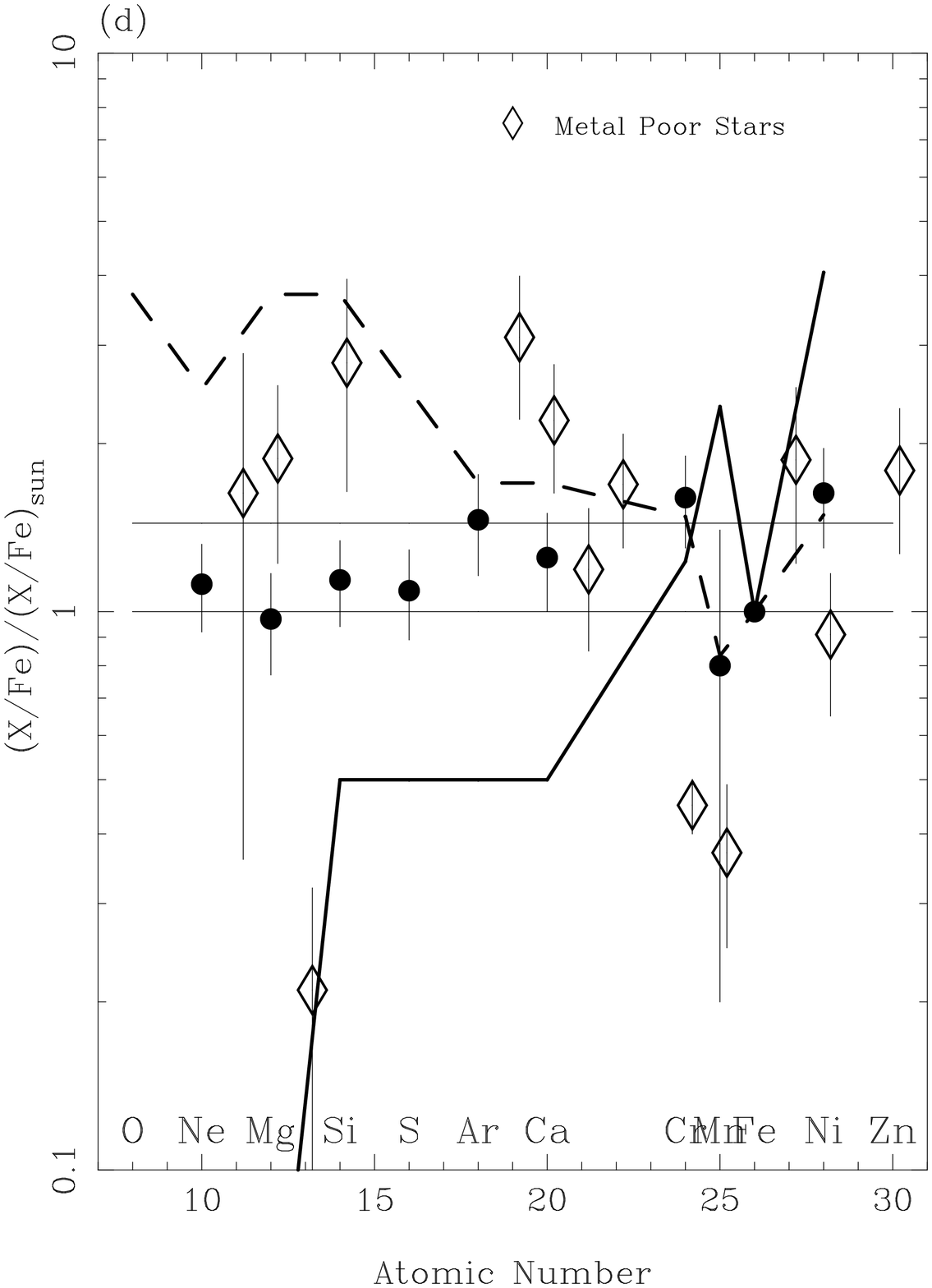}
\end{center}
\caption{(a) Elemental abundance ratios determined from the two temperature fits.
X/Fe (X=Ne, Mg, Si, S, Ar, Ca, and Ni) are shown in relative to those of the solar value 
(i.e. (X/Fe)/(X/Fe)$_\sun$).
Typical errors are given for the $2'-4'$ region.
The Fe abundance is calculated by averaging over those of the two components.
(b) 
Relative differences of the best fit parameters.
Filled circles show differences between best-fit values
obtained from the FI fit and those from the combined one (FI+BI) normalized by the latter value.
Open circles show those from the BI fit.
Points connected with solid- and dashed-lines are from the $0'.5-2'$ and $2'-4'$ regions, respectively.
Statistical errors are given only at the BI $0'.5-2'$ values.
Compared with these errors, errors of other values are similar or small.
(c) Comparison with previous X-ray measurements. 
Results of the Fornax cluster \citep{matsushita03},
A~1060 \citep{sato07}, average values of 22 clusters \citep{dePlaa07} 
and M87 \citep{matsushita03} are shown along with the present result of the Perseus cluster ($2'-4'$ region).
The two vertical lines indicate 'photospheric' (1.0) and 'meteoritic' (1.44) solar abundances, 
corresponding to iron number density, Fe/H, of $4.68\times10^{-5}$ and $3.24\times10^{-5}$, respectively.
(d) 
Same for (c), but with abundances in extremely metal poor stars in the Galaxy \citep{cayrel04}. 
Predictions from type Ia and type II SNe are also shown in solid- and dashed-lines, respectively.
Model yields are taken from \citet{iwamoto99} (W7 model for the type Ia).
\label{fig:abun}
}
\end{figure*}

The fitting result is given in Table~\ref{tbl:2t-fit}.
Statistical errors of elemental abundances are $5-10$\% of the best fit value.
% note 2-p.121, 
% ~/KLM/suzaku/perseus/xis/0423-1/2d-2VapecPow/wi3-2t132-fi.log
Obtained abundances are shown in Fig.~\ref{fig:abun}(a).
Here we use the Fe abundance averaged over those of the two components by weighting by the emission measure.

\begin{figure*}
\begin{center}
\includegraphics[angle=-90,scale=.43]{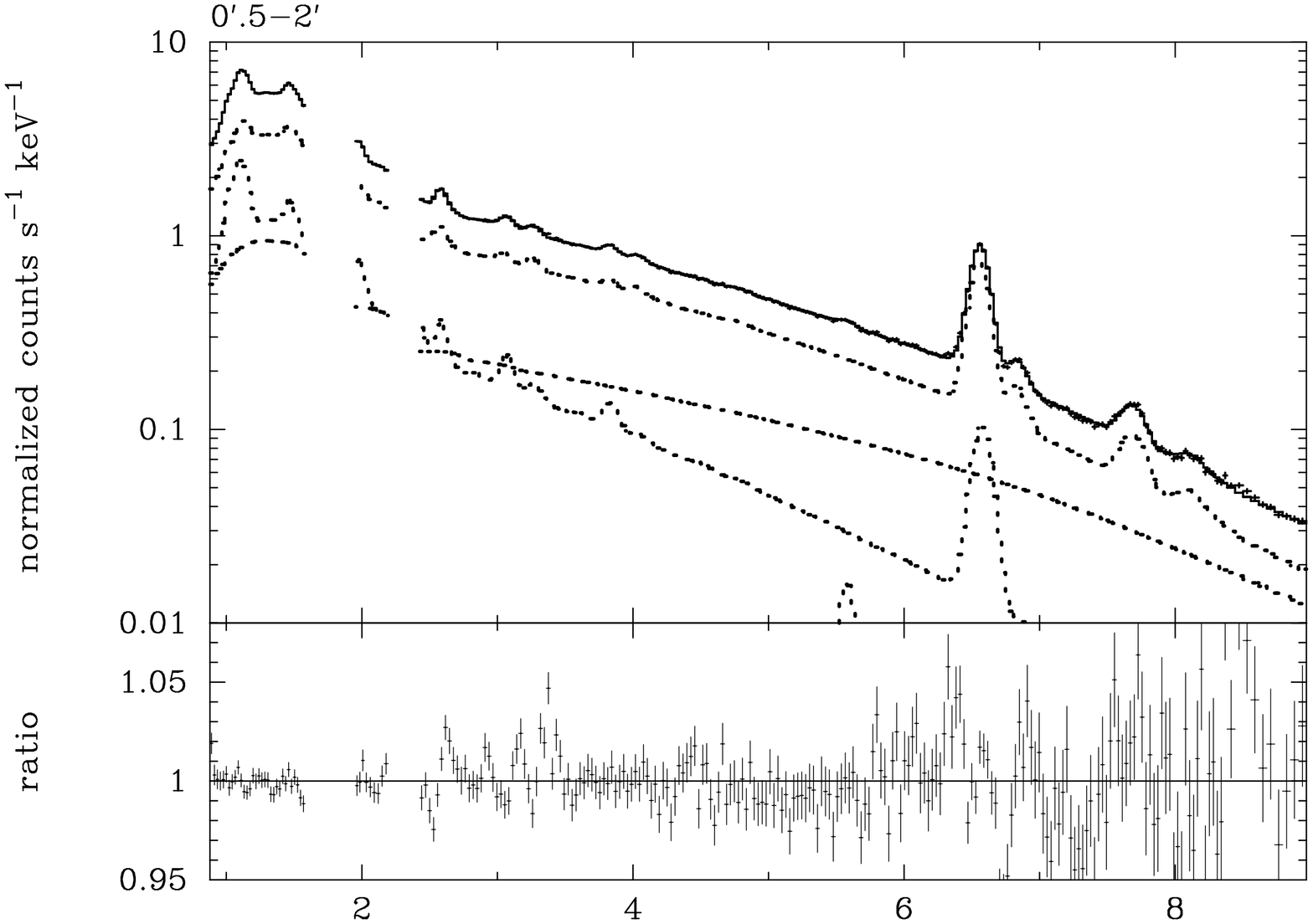}
\includegraphics[angle=-90,scale=.43]{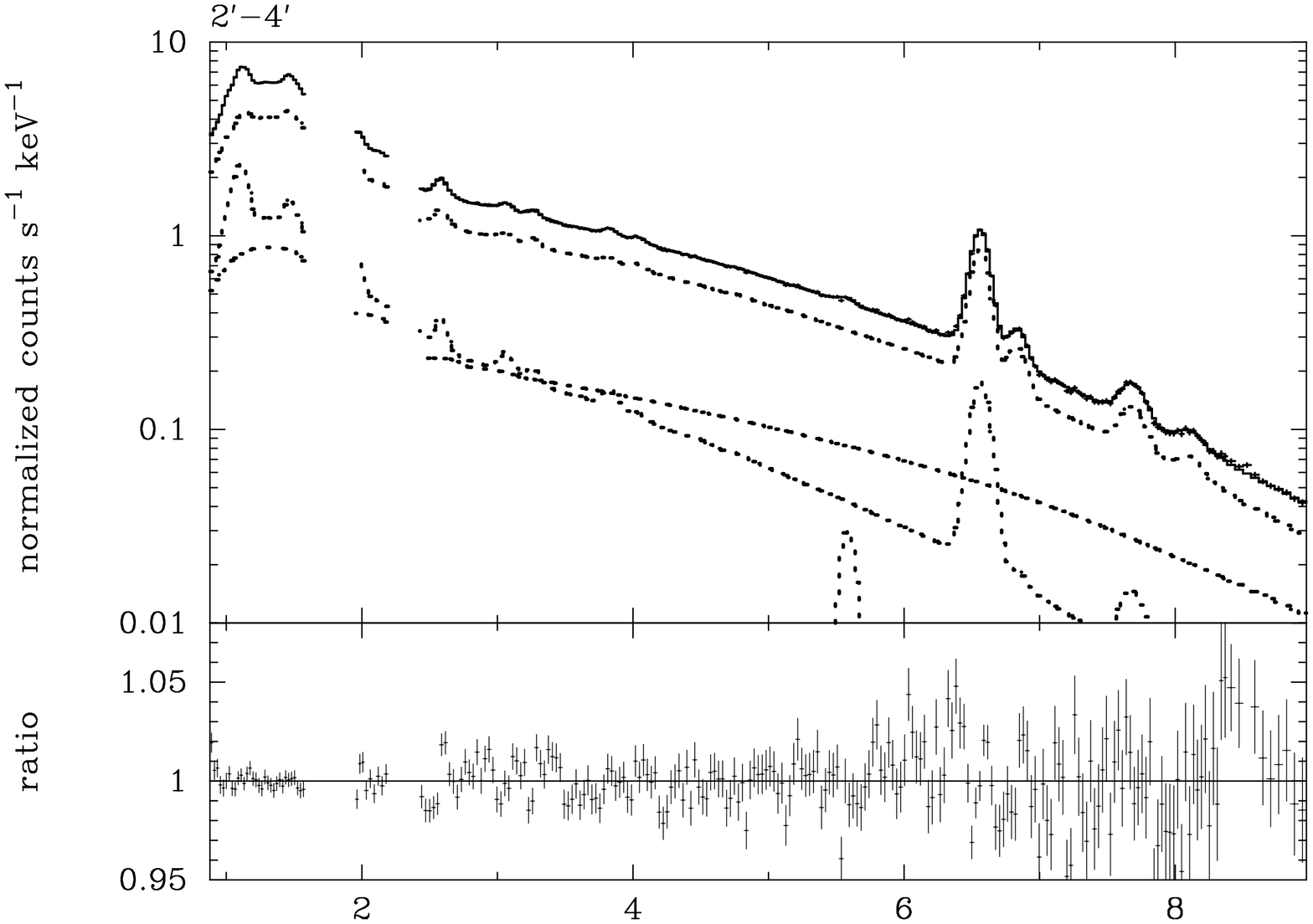}
\includegraphics[angle=-90,scale=.43]{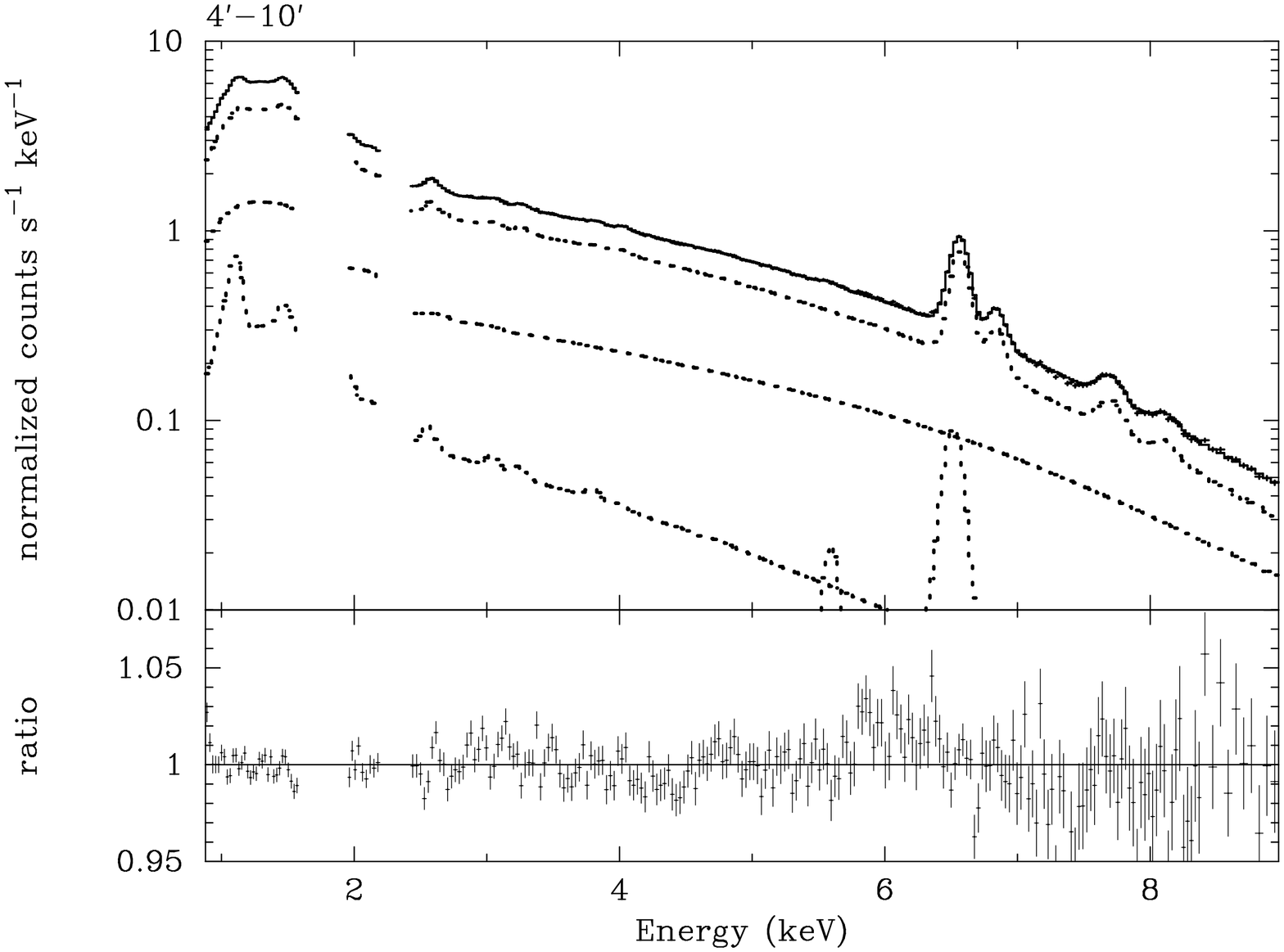}
\end{center}
\caption{Cluster spectra along with the best-fit model.
From top to bottom, the FI-CCD data from $0.5'-2'-4'-10'$ regions are shown.
The model components are shown by lines in upper panels.
In lower panels fit residuals in terms of the data to model ratio are shown.
\label{fig:2t-fit}
}
\end{figure*}

Examples of the spectral fitting are shown in Fig.~\ref{fig:2t-fit}.
There are two types of systematic residuals.
The one is those found around strong lines, 
such as the Fe K line at $\sim$ 6.5~keV in the $0'.5-2'$ spectral fit.
Limited accuracies in the energy response
could cause these discrepancies.
Furthermore, uncertainties in the modeling of the Fe K line structure could be larger than the statistics of the current data
\footnote{To assess possible uncertainties in the Fe K line modeling, 
we compared the APEC with the MEKAL model implemented in the XSPEC. 
After convolving models with the XIS response, 
we found differences between the two of $5-10$\%. 
A similar level of model uncertainties is expected.}.
The other residual can be seen in the energy range of $5.5-7.5$~keV only in the $4'.0-10'$ fit.
This could be related to instrumental background.
At this energy band, the background is dominated by line emission such as 
Mn K$_\alpha$ (5.9~keV), 
Mn K$_\beta$ (6.5~keV), and Ni K$_\alpha$ (7.5~keV).
Time variations in line fluxes and changes in the line spread function
create errors on the background estimation \citep{tawa08}.
Nevertheless, each model describes the data with residuals smaller than five percent in most energy bins.

\begin{deluxetable}{llllllllll}
\tabletypesize{\scriptsize}
%\rotate
\tablecaption{The fitting results using the two temperature components.
\label{tbl:2t-fit}
}
\tablewidth{0pt}
\tablehead{
\colhead{Region } &\colhead{XIS} &\colhead{$\chi^2/d.o.f.$ } &
\colhead{T1} & \colhead{Fe1} & \colhead{ Norm1$\tablenotemark{a}$} & 
\colhead{T2} & \colhead{Fe2} & \colhead{ Norm2$\tablenotemark{a}$} & 
\colhead{P-Norm$\tablenotemark{b}$} \\
\colhead{(radii)} & \colhead{} & \colhead{}& \colhead{(keV)} & \colhead{(solar)} & \colhead{}& \colhead{(keV)} & \colhead{(solar)} & \colhead{} 
}
\startdata
$0'.5-2'$ & FI+BI & 1246/471 & $4.2\pm0.04$ & $0.62\pm0.01$ & $0.37\pm0.01$ & $2.2\pm0.1$ & $0.71\pm0.02$ & $0.13\pm0.01$ & $3.1\pm0.1$\\
$2'-4$ & FI+BI & 1207/471 & $5.0\pm0.1$ & $0.53\pm0.01$ & $0.39\pm0.01$ & $2.5\pm0.05$ & $0.83\pm0.04$ & $0.12\pm 0.01$ & $2.5\pm0.1$\\
$4'-10'$ & FI & 485/239 & $5.7\pm0.1$  & $0.45\pm0.01$ & $0.56\pm0.01$ & $2.9\pm0.1$ & $1.5\pm0.5$ & $0.036\pm0.01$ & $4.9\pm0.2$\\
\enddata
\tablenotetext{a}{Normalization of VAPEC component in the unit used in XSPEC.}
\tablenotetext{b}{Normalization of the power-law component in unit of $10^{-2}$ photons keV$^{-1}$ s$^{-1}$ at 1~keV.}
\end{deluxetable}

%%% 
To estimate systematic errors associated with the instrumental calibration, 
we fit the FI and BI spectra separately 
and compare the results with those obtained from the combined (FI and BI) fits above.
In these fits, statistical errors on abundances are
about $5-10$\% and $10-15$\% for the FI and BI spectra, respectively. 
As shown in Fig.~\ref{fig:abun}(b) relative differences between the separate and combined fits are within 20\% mostly and within 30\% in all cases. 
By this comparison and the current calibration status of the XIS\footnote{See ``The Suzaku Technical Description'' available at \url{http://www.astro.isas.jaxa.jp/suzaku/doc/suzaku\_td/}.},
we suppose that the possible systematic errors on the abundances are about $\pm (20-30)$\% of the best-fit values.

%(resonant scattering)\\
In the core of the Perseus cluster,
resonant scattering may redistribute some Fe line photons spatially from the core to the outer region.
We examined this effect by following \citet{churazov04}.
The resonant scattering should be relatively significant on the resonance transition of He-like Fe (1s$^2$-1s2p) at 6.70~keV.
Then, we exclude this line energy band ($6.3-6.8$~keV) and fit the data again.
This fit gives consistent abundances of Fe and Ni with the previous one.
This indicates that the  scattering is insignificant in our data.
Using \xmm data, \citet{churazov04} and \citet{gastal04} independently reported consistent results and
suggested that the scattering effect is reduced because of gas motions in the core.

\subsection{Cr and Mn Line emission}
\begin{figure*}
\begin{center}
\includegraphics[scale=.40]{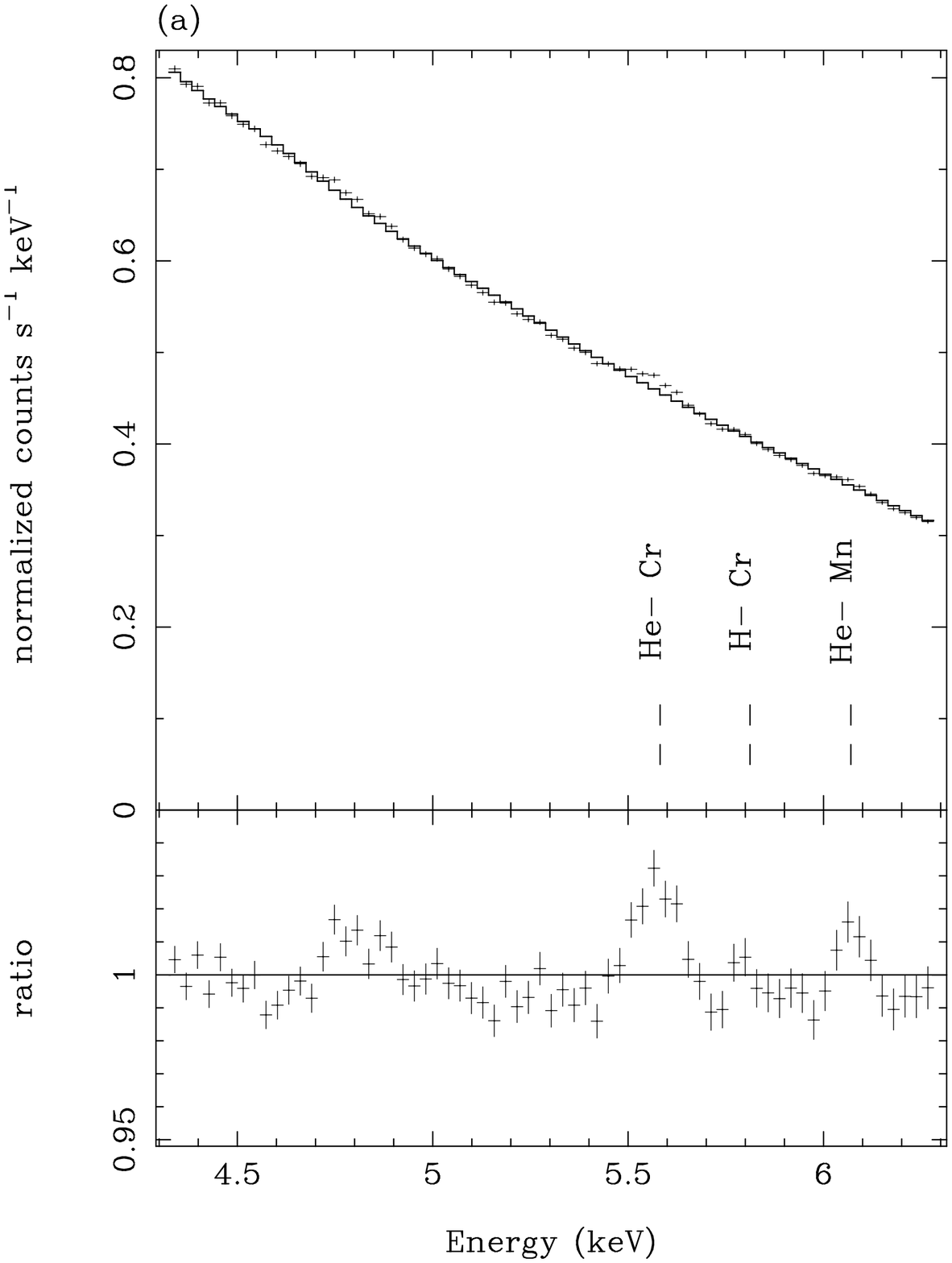}
\includegraphics[scale=.40]{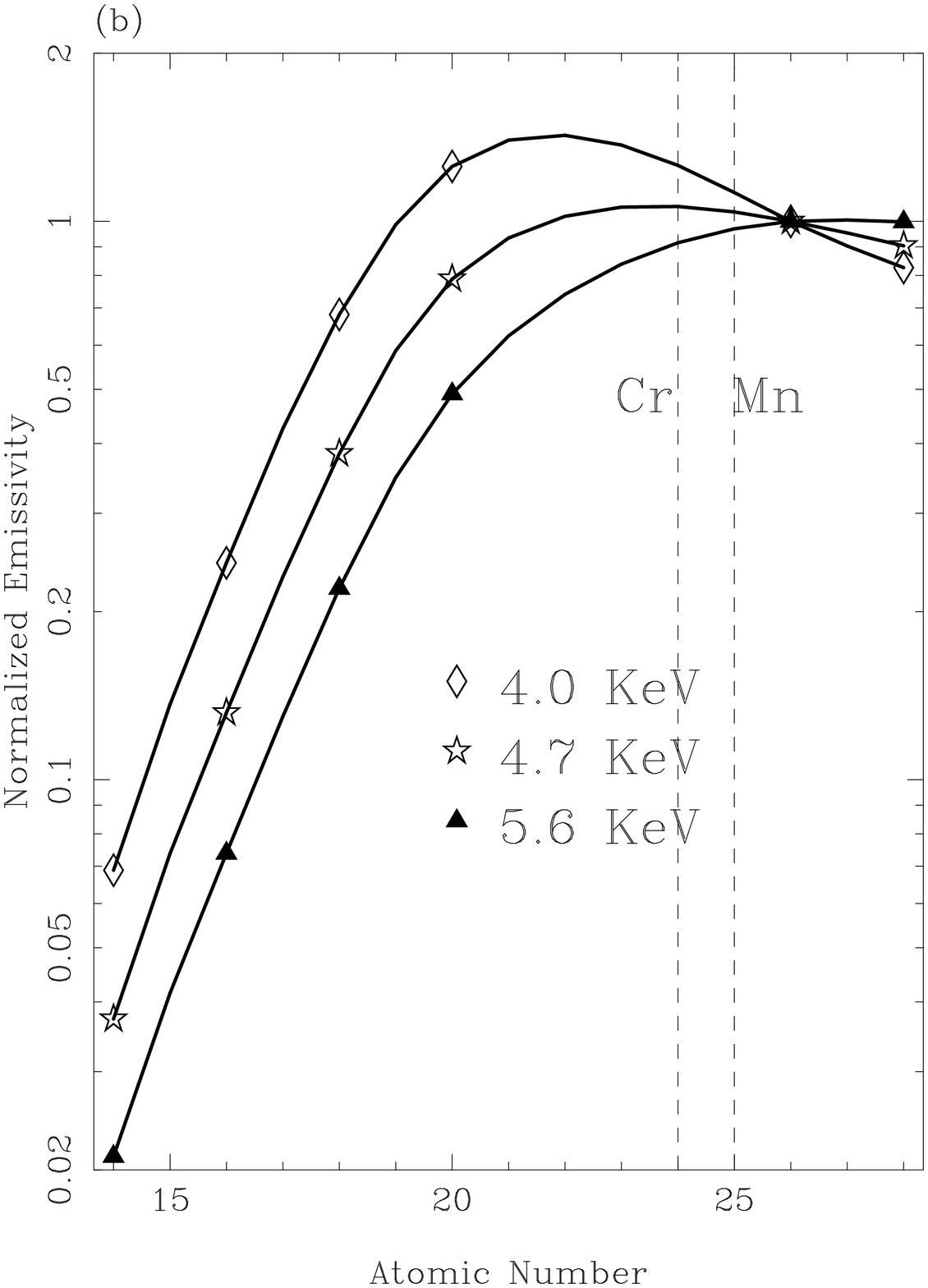}
\end{center}
\caption{(a): 
Spectrum around the Cr and Mn lines extracted from the $0'.5-10'$ region.
The FI CCD data with and without spaced-row charge injection are combined.
The positions of He- and/or H-like Cr and Mn are indicated.
(b): X-ray line emissivities from He-like ions as a function of the atomic number.
The model values for Si, S, Ar, Ca, Fe, and Ni at three different temperatures
obtained from the ATOMDB are shown by various mark points.
Lines are spline fit to these values.
Positions of Cr and Mn are indicated  by dashed vertical lines.
\label{fig:Cr}
}
\end{figure*}

As mentioned above, we detected X-ray line emission from Cr.
Next to Cr, we expect line emission from Mn from a sub-solar metallicity plasma.
Note that the Cr and Mn solar abundances are about 1\% and 0.5\% of that of Fe, respectively.
To search for these lines, we use the FI spectra integrated over a $0'.5-10'$ region in spaced-row charge injection {\it on} and {\it off} modes together.
As shown in Fig.~\ref{fig:Cr}(a), by comparing the data with a bremsstrahlung continuum model, we have detected a line from He-like Cr clearly and that from He-like Mn. 
Note that in the model used in this fit the Si-escape line associated with the He-like Fe K$_\alpha$ line
is included. This Si-escape line has a peak at 4.8~keV with a flux of about 1\% of the data at that position.

Using the same data above and Gaussian line fitting, we measured these line positions and fluxes as given in Table~\ref{tbl:cr-mn}.
The detection of the Cr line is more than 99.9\% confidence level, 
while that of the Mn line is more than 98\% level.
The fluxes are measured with respect to that of the He-like Fe K$_\alpha$ line at 6.70~keV.
The resonance transitions of He-like Cr and Mn are at 5.682~keV and 6.181~keV, respectively \citep{hwang00}.
The obtained line positions are fully consistent with those of redshifted resonance positions.
The equivalent widths of the Cr and Mn lines are about 5~eV and 2~eV, respectively.

In a similar way as above, we examined spectra from the three regions separately.
The significances of the Cr line detections are $>99.9$\%, $>99.9$\%, and 94\% confidence 
from the $0'.5-2'-4'-10'$ spectra, respectively.
The Mn line detection is significant ($>90$\% confidence) only from the $0'.5-2'$ spectrum.
In Table~\ref{tbl:cr-mn}, we show the obtained line positions and fluxes.

To check if these features are instrumental,
we examine the Crab nebula and instrumental background data obtained with the XIS.
The Crab has a bright power-law X-ray emission and is a standard calibration source for X-ray instruments.
In the Crab spectra, there is no structure at the line positions of Cr and Mn detected above.
We found no structure at the same positions in the instrumental background.
There are lines from neutral Mn K$_{\alpha}$ (5.895~keV) and K$_{\beta}$ (6.490~keV).
However, these position are away from those of He-like Cr or Mn.
Moreover, we confirmed that the \chandra deep spectrum ($\sim $1~Ms exposure) shows a line feature at the same Cr position with a flux consistent with the \suzaku data.
These facts, redshifted line positions, and the obtained flux as expected from 
a sub-solar metallicity plasma (as calculated in discussion) indicate that we detected the Cr and Mn line emission from the cluster.
 
In our knowledge, this is the first significant detection of X-ray line emission of Cr and Mn ions from an extragalactic source.
\citet{werner06} reported the detection of the same Cr line from the cluster 2A~0335+096 at the $2\sigma$ level. 

\begin{deluxetable}{llllllll}
\tabletypesize{\scriptsize}
%\rotate
\tablecaption{The Cr and Mn parameters obtained from the spectral fitting.
\label{tbl:cr-mn}
}
\tablewidth{0pt}
\tablehead{
\colhead{Region}  & 
\colhead{Cr } & \colhead{Cr/Fe} & 
\colhead{Mn } & \colhead{Mn/Fe} & 
\colhead{Cr/Fe } & \colhead{Mn/Fe}
& \colhead{$kT$ }\\
\colhead{}  & 
\colhead{position (eV) } & \colhead{flux $\tablenotemark{a}$} & 
\colhead{position (eV) } & \colhead{flux $\tablenotemark{a}$} & 
\colhead{ratio (solar) $\tablenotemark{b}$} & \colhead{ratio (solar) $\tablenotemark{b}$}
& \colhead{(keV)$\tablenotemark{c}$}
}
\startdata
$0'.5-10'$ & $5574\pm 10$ & $1.6\pm0.3$ & 6070$\pm30$ & $0.4\pm0.2$ & -- & -- & --\\
$0'.5-2'$ & $5571\pm 35$ & $1.81\pm0.35$ & 6070 & $0.49\pm 0.35$ & $1.4\pm0.3$ & $0.8\pm0.6$ & 4.0 \\
%% see p.104 T=4keV, 
$2'-4'$   & $5582\pm 30$ & $1.76\pm0.35$ & 6070& $0.41\pm 0.32$ & $1.6\pm0.3$ & $0.8\pm0.6 $ & 4.7\\
$4'-10'$  & $5573\pm 40$ & $1.53\pm0.46$ & 6070 & $0.4(<0.9)$   & $1.7\pm0.5$ & $0.8 (<1.8)$ & 5.6\\
\enddata
\tablenotetext{a}{Flux ratio to that of the He-like Fe K$_\alpha$ line at 6.7~keV in units of $10^{-2}$.}
\tablenotetext{b}{Elemental abundance ratio relative to the solar value.}
\tablenotetext{c}{Assumed temperature.}
\end{deluxetable}

\section{SUMMARY AND DISCUSSION}
\suzaku XIS observations of the Perseus cluster provided one of the best quality X-ray spectra from clusters.
We have detected the He-like Cr and Mn lines.
Furthermore, we measured the radial distribution of abundances of Ne, Mg, Si, S, Ar, Ca, Fe, and Ni.
There is no strong radial variation in relative abundance ratios such as Mg/Fe and Si/Fe.
The abundance ratios are within a range of $0.8-1.5$ times the solar value.

\subsection{Cr and Mn abundances}
Based on observed line fluxes of Cr and Mn ions, we estimate abundances of these elements.
There is no atomic data in the ATOMDB \citep{smith01} nor MEKAL \citep{kaastra92} providing X-ray emissivities of these lines.
Then, following \citet{hwang00}, we use line emissivities of elements (Si, S, Ar, Ca, Fe, and Ni) available from the ATOMDB 
and estimate those of Cr and Mn.
Here we assume that emissivities per single ion of these elements at a given temperature 
is a smooth function of the atomic number.
We interpolate the function using spline fits, as shown in Fig.~\ref{fig:Cr}(b).
We used temperatures (Table~\ref{tbl:cr-mn}) calculated by averaging over the two thermal components in the best-fit model obtained in subsection 3.2. 
Errors introduced by the interpolation would be smaller than $10-20$\%.
This estimation along with observed fluxes gives Cr and Mn abundances (Table~\ref{tbl:cr-mn}): 
Cr/Fe and Mn/Fe ratios are $1.6\pm0.4$ and $0.8\pm 0.6$ times the solar, respectively.

The above estimated values obtained for Cr and Mn along with the Fe and Ni abundances derived in the spectral fitting 
provide iron-group abundances for the first time in the ICM.
We found Cr:Mn:Fe:Ni ratios close to the solar value.
These abundance patterns depend on the nature of their origins, supernovae (SNe).
For example, \citet{badenes09} propose to reveal the detailed nature of type Ia SN using Mn, Cr, and Fe X-ray lines in SN remnants.
In the case of the ICM metals, 
we can examine SNe averaged over many events in member galaxies.
The present result 
supports the idea that the metals in the ICM and solar neighborhood were produced
in a similar way. 

\subsection{Abundance Pattern}
Our derived abundance pattern is compared with relevant observations and theories below.
Some comparisons are shown in  Fig.~\ref{fig:abun}.

First, the \chandra result in the Perseus cluster \citep{sanders04} is compared.
We used their abundance pattern obtained from a region (40-80~kpc) similar to our data,
but accounted for projection. 
The Ne/Fe, Si/Fe, Ar/Fe, Ca/Fe, and Ni/Fe ratios are consistent with the \suzaku result.
In contrast, the \chandra data shows lower Mg/Fe ($\sim 0.5$) and higher S/Fe ($1.5-1.8$) values 
than the solar ratio found in the current study.
We suppose that these inconsistencies are caused partly by calibration errors used 
in \citet{sanders04} and partly caused by the difference in the spectral extraction method.
Indeed, we found that much longer \chandra data ($\sim$ 1~Ms exposure)
with the latest calibration and without the projection correction give
Mg and S abundances consistent with our result.

Second, we compared our result with measurements in other clusters [Fig.~\ref{fig:abun}(c)].
Note that including our study, most results are derived from cluster central regions (within a few 100~kpc).
In most elements, our results are consistent with those in other clusters.
Exceptions are lower Ar and higher Ca abundances in average values over a sample of 22 clusters
reported by \citet{dePlaa07}.
We cannot find this trend in the Perseus cluster.
For all measured elements from Ne to Fe, the cluster abundances are consistent with the solar ratios within uncertainties.
The Ni/Fe ratios are relatively high.
This may be a sign of a contribution from type Ia SNe \citep[e.g.][]{dupke00}.
The sub-solar value of O/Fe measured in other studies may have a similar origin \citep[e.g.][]{tamura01}.

Third,
our measurement is compared 
with that from Galactic stars with extremely low metallicity [Fig.~\ref{fig:abun}(c); \citet{cayrel04}]. 
The latter abundances should be close to those originated from the first generation of stars and SNe in the universe.
The patterns are very different.
Among elements that we measured, 
elements lighter than Cr are more abundant in the metal poor stars, 
while Cr, Mn, and Ni are more abundant in the ICM.
This comparison, 
regardless of theoretical models, 
indicates that the ICM is polluted not solely by the first generations of stars
but also by additional source(s) significantly.
The additional source(s) should add iron group elements to the ICM.
The most-likely dominant contributor is type Ia SNe,  
as in the Galactic chemical evolution model \citep[e.g.][]{kobayashi06}.
The abundance pattern in the Perseus  cluster is closer to the solar value than to those in extremely low metallicity stars.
Therefore the contribution of the first generations of stars into the ICM metal is not required largely, 
in contrast to the model proposed in \citet{loe01}. 

Finally, the measured pattern is compared with theoretical predictions from two types of SNe [Fig~\ref{fig:abun}(d)].
The observed abundance pattern cannot be produced solely by each type of SNe.
All the observed abundance ratios are between the two predictions, 
except for Cr and Mn, which have relatively large uncertainties.
Consequently, we conclude that the metals in the ICM of the Perseus and other clusters were produced by a mix of both types of SNe.
Similar conclusions have been obtained from \asca, \chandra, and \xmm studies
\citep[e.g][]{fuka98, tamura01, sanders06}.
The present data strengthen this scenario.

\acknowledgments
We thank J. S. Kaastra, K. Hayashida, and anonymous referee for their comments.

\end{document}